\documentclass[conference,10pt]{IEEEtran}
\usepackage{cite}

\usepackage{graphicx}
\usepackage{ifpdf}
\ifpdf
  \usepackage{epstopdf}
\fi
\usepackage[cmex10]{amsmath}
\usepackage{amsthm}
\usepackage{amssymb}
\usepackage{cases}
\usepackage{bm}

\usepackage[caption=false,font=footnotesize]{subfig}
\usepackage{fixltx2e}
\usepackage{url}

\begin{document}
%
\title{Graph-based Framework for Flexible Baseband\\ Function Splitting and Placement in C-RAN}
%
%
%

\author{\IEEEauthorblockN{Jingchu Liu, Sheng Zhou, Jie Gong, Zhisheng Niu}
        \IEEEauthorblockA{Tsinghua National Laboratory for Information Science and Technology\\
        Department of Electronic Engineering, Tsinghua University\\
        Beijing 100084, China \\
        Email: liu-jc12@mails.tsinghua.edu.cn,  sheng.zhou@tsinghua.edu.cn\\ gongj13@mail.tsinghua.edu.cn, niuzhs@tsinghua.edu.cn\\}
        \and
        \IEEEauthorblockN{Shugong Xu}
        \IEEEauthorblockA{Intel Labs\\
        \\
        Beijing 100080, China\\
        Email: shugong.xu@intel.com}
        }

%
%



\maketitle

\begin{abstract}
The baseband-up centralization architecture of radio access networks (C-RAN) has recently been proposed to support efficient cooperative communications and reduce deployment and operational costs. However, the massive fronthaul bandwidth required to aggregate baseband samples from remote radio heads (RRHs) to the central office incurs huge fronthauling cost, and existing baseband compression algorithms can hardly solve this issue. In this paper, we propose a graph-based framework to effectively reduce fronthauling cost through properly splitting and placing baseband processing functions in the network. Baseband transceiver structures are represented with directed graphs, in which nodes correspond to baseband functions, and edges to the information flows between functions. By mapping graph weighs to computational and fronthauling costs, we transform the problem of finding the optimum location to place some baseband functions into the problem of finding the optimum clustering scheme for graph nodes. We then solve this problem using a genetic algorithm with customized fitness function and mutation module. Simulation results show that proper splitting and placement schemes can significantly reduce fronthauling cost at the expense of increased computational cost. We also find that cooperative processing structures and stringent delay requirements will increase the possibility of centralized placement.
\end{abstract}


%
\IEEEpeerreviewmaketitle

\section{Introduction}
Recently, the number of smart devices has grown into billions, and the wide collection of mobile applications is increasingly interleaved with our daily lives. As a result, next generation wireless communication systems have received unprecedented expectations, aiming at $1000$ times capacity, $100$ times data rate, billions of devices, and millisecond-level delay. To fulfill these goals, the solution envisioned so far is a dense, cooperative\cite{CHORUS}, and heterogeneous \cite{HCN} wireless network. However, realizing such a network with the traditional, distributed architecture for radio access network (RAN) means high capital expenditure ({CAPEX}) and operational expenditure ({OPEX}); and cooperative communications will be highly limited due to the bandwidth bottleneck between distributed base stations.

The baseband-up centralization architecture, e.g. {C-RAN}\cite{cran}, is a promising solution to these problems. In this architecture, radio signals are first digitized at the antenna sites by remote radio heads ({RRHs}), and then transported via the so called fronthaul network to the centralized baseband processing units ({BBUs}) for signal processing. The benefits of such an architecture include reduced {CAPEX} and {OPEX}\cite{cran,vbsmodel}, efficient information exchange for cooperative communications\cite{cran}, and increased flexibility due to the use of general purpose platforms {(GPPs)}\cite{cloudiq}. Despite these benefits, a major challenge for baseband-up centralization is the huge aggregation bandwidth requirement for fronthaul network. For a typical long-term evolution ({LTE}) cell configuration of $20$ MHz wireless bandwidth and $8$ antennas, $10$ Gbps fronthaul bandwidth is required in downlink ({DL}) or uplink {(UL)} to transport baseband samples\cite{cpri}. The demand for fronthaul bandwidth would only be higher with even larger wireless bandwidth and more antennas.

To cope with this problem, a number of baseband compression algorithms have been proposed. Time-domain compression algorithms \cite{bbTWC} are 
simple and fast, but they only provide limited compression performance ($2-3 \times$). Another kind of algorithms perform compression in frequency domain, which can achieve $20 \times$ compression rate.
However, frequency-domain methods requires a great amount of computation to perform FFT/IFFT and thus may suffer from long delay.

Lorca and Cucala \cite{lorca13} propose a novel method that can significantly reduce {DL} fronthaul bandwidth ($30 \times$) by relocating modulation and precoding processing functions from the central office back to remote sites. The intuition behind this method can be explained with some insights into baseband processing. In a sense, {DL} baseband processing functions are designed to add artificial redundancies into communication signals in order to combat the impairments of wireless channels\footnote{For example, modulation turns constellation codewords, which can be represented with only a few bits ($6$ bits for $64$ QAM), into complex constellation samples, which is usually digitized with tens of bits ($30$ bits for {LTE}). Other baseband functions like channel encoding and beamforming, also introduces other types of artificial redundancies.}. Because redundancies are accumulated function-by-function along the processing chain, baseband-up centralization actually needs to transport the most redundant signal. In constrast, the method in \cite{lorca13} no longer needs to transport the redundancies introduced by modulation and precoding, therefore fronthaul bandwidth can be reduced.

Nevertheless, the above method does not consider the cost for accommodating additional baseband functions (computational cost) at remote sites. In reality, computational cost can become a major constraint at remote sites due to power consumption and form factor considerations. Also, some wireless protocols have stringent real-time requirements for processing baseband tasks\footnote{In {LTE}, the processing of a subframe should be completed within $3$ ms for timely hybrid automatic retransmission request (HARQ).}. Hence, the influence on processing delay should also be considered. For these reason, we propose in \cite{CONCERT} to flexibly split and place baseband functions in the network based on the cost profile and delay requirements of different applications. But an analytical framework for deciding the optimum splitting and placement scheme is still needed.

\begin{figure}[!t]
	\centering
	\includegraphics[width=3in]{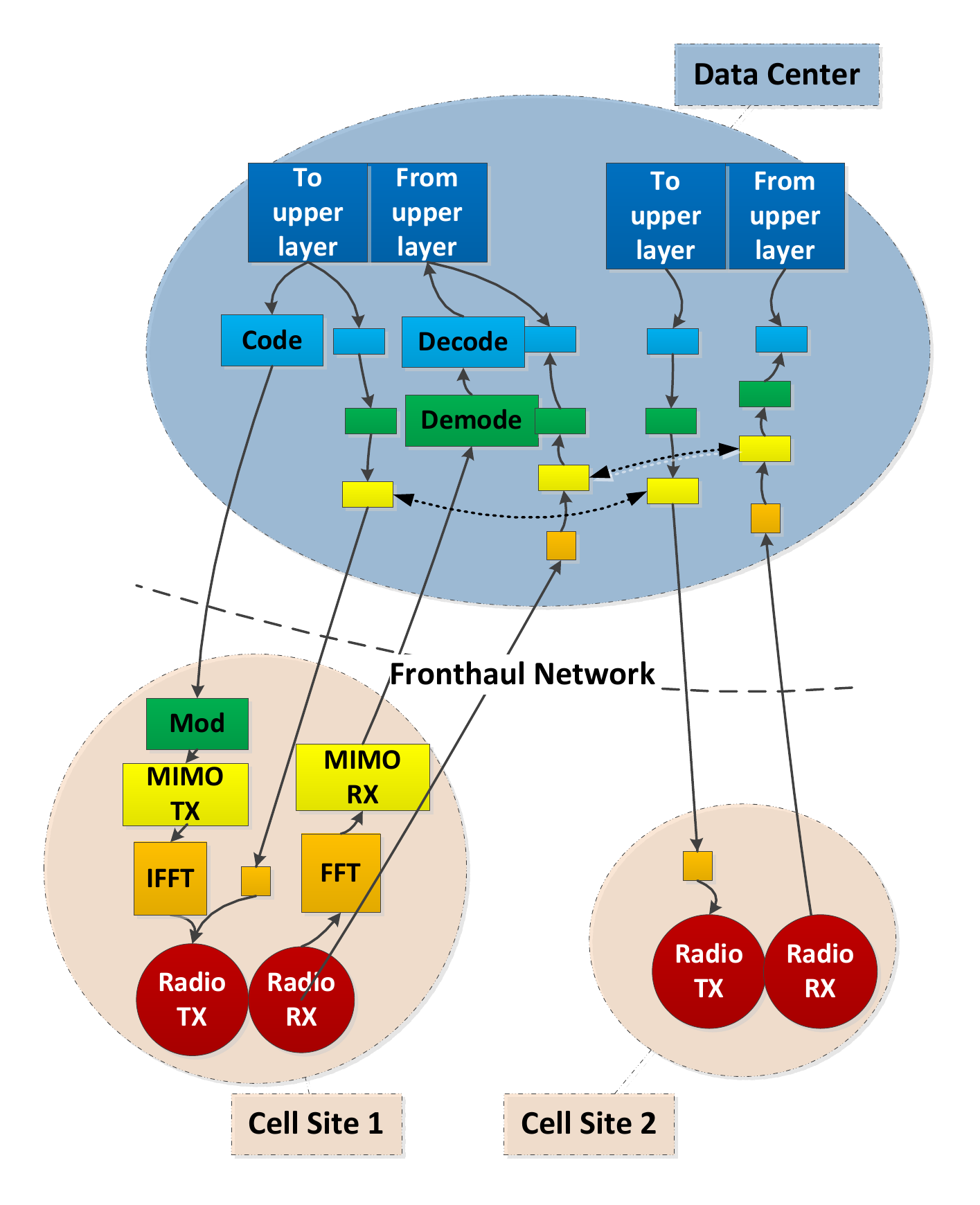}
	\caption{Flexible splitting and placement of baseband functions.}
	\label{graph}
\end{figure}

In this paper, we present a graph-based framework for baseband function splitting and placement. We first translate baseband transceiver structures into directed graphs so that the splitting and placement problems can be formulated as graph-clustering problems (as illustrated in Fig. \ref{graph}). We then propose a genetic algorithm with customized fitness function and mutation module for the graph-clustering problem. Simulation results show that the proposed algorithm can effectively reduce fronthauling cost. An anatomy of these results also reveals that cooperative structures and stringent delay constraints will result in more centralized function placement.

The rest of the paper is organized as follows. In section \ref{model}, we represent baseband processing structures using directed graph and formulate the baseband function splitting and placement problem as a graph-clustering problem. In section \ref{algorithm}, we introduce the proposed customized genetic algorithm. Simulation results are presented and discussed in section \ref{simulation} and the paper is concluded in section \ref{conclusion}.

\section{Model Formulation}\label{model}
In this section, we present the proposed graph-based framework. For better understanding, we also present a concrete example for mapping baseband function splitting problems into graph-clustering problems.
\subsection{Baseband processing structures and directed graphs}
We represent a baseband processing structure with a directed graph $G=(V,E)$. Each node $v \in V$ stands for an atomic baseband processing function such as FFT or MIMO detection, and each \emph{directed} link $e \in E$ represent the logical connectivity between the nodes it connects \footnote{We assume nodes and edges are indexed using with integer values.}. Each node is assigned with a node weight according to the node complexity function $\gamma$ : $V \rightarrow \mathbb{R}$, which indicates the computational complexity of this processing node. And each link is assigned with a link weight by the link bandwidth function $\omega$ : $E \rightarrow \mathbb{R}$, which represents the amount of information that has to be exchanged between the processing node it connects. Note some of the nodes are sources (no inbound links) or sinks (no outbound links). Each distinct path $p \in P$ from a source to a sink represents a complete chain of baseband processing functions\footnote{There may be multiple paths between a pair of source and sink due to the parallel processing of channels for different users.}.

The graph formulated in this way may contain cycles. The reason is that there may be mutual information exchange between baseband functions in cooperative systems such as cooperative multipoint ({CoMP}) processing or multi-user MIMO. These cycles are important features of the whole system and have significant influence on the choice of splitting and placement. However, we do assert that there are no self-cycles, which may appear due to inappropriate abstraction of iterative processing function. The information flow of iterative processing functions should be embedded in the atomic baseband processing functions to avoid self-cycles.

\subsection{Function Splitting and graph clustering}
With this representation, we can express function splitting and placement as graph clustering schemes $\xi$: $V \rightarrow \mathbb{Z}$, which assign nodes to a collection of clusters. Note that in our model, clusters have explicit physical meanings. Different clusters correspond to different physical locations (e.g. remote sites and central office), and the nodes in the same cluster correspond to baseband processing functions that are placed at the same physical location. The links between clusters correspond to the information flow to be transported by the fronthaul network.  

The study of graph clustering is concerned with grouping nodes in order to optimize some cost/gain metric. The classic goal is to group nodes that are ``close to'' or ``similar to'' each other \cite{graphclustering}. We employ different goals in our formulation to address the special concerns of baseband function splitting and placement. The goals reflect the computational cost for accommodating nodes at some location and the fronthauling cost for transporting data between different locations using fronthaul networks.

Specifically, we define a pair of cost metrics. The first metric is computational cost $c_c(i;\xi)$, where $i$ is the index of a cluster, and $\xi$ is the clustering scheme under consideration. The computational cost is to reflect the cost of implementing baseband processing functions at a physical location and should thus be a function of the total node complexity inside the cluster. Also, because computational cost often differs in different locations in real world\footnote{For example, it is more expensive to accommodate computation at remote sites than at central office due to form factor, electricity, and site rental costs.}, clusters are allowed to have different cost profiles. The second metric is fronthauling cost $c_f(i,j;\xi)$, where $i$ and $j$ are the index of two clusters, and $\xi$ is the clustering scheme. Fronthauling cost is to reflect the bandwidth required to transport information between different physical locations. As a result, it should be a function of the total edge weights between clusters. As we have explained from the perspective of redundacy, computational cost and fronthauling cost are in generals contradiction goals to optimize. Thus, different clustering scheme will result in different tradeoffs between computational and fronthauling costs. For this reason, we aim to characterize the tradeoff between computational cost and fronthauling cost.

Another important feature of our model is the path delay constraint, which is imposed to guarantee the real-time processing of communication signals. We assume that each node on a path will impose an additional delay $d(v,p;\xi)$ to this path, where $v$ is the index of the node and $p$ is the path under consideration. This delay function captures the processing and buffering latency of baseband tasks. Any valid clustering scheme should guarantee that the total delay of a path is smaller than a predefined threshold $D(p)$: $d(p;\xi)=\sum_{v\in p}{d(v,p;\xi)}<D(p).$ 

\subsection{Example}\label{Example}
\begin{figure}[!t]
	\centering
	\includegraphics[width=3.7in]{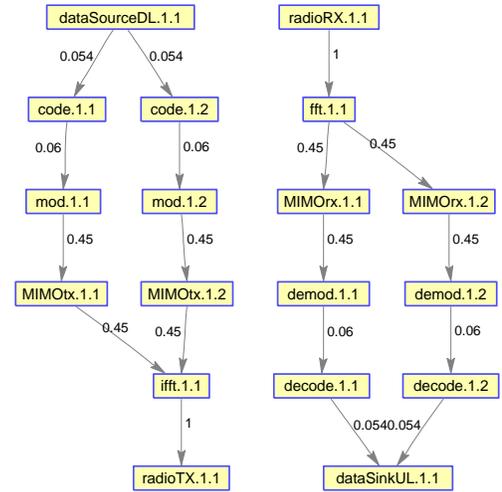}
	\caption{A simplified baseband processing architecture }
	\label{baseband1}
\end{figure}

The baseband processing structure used in our simulation is shown in Fig. \ref{baseband1}. We use two such baseband processing structures to represent  two cells. Note this structure is just a simplification of real-life physical layer baseband structure. We only include some of the most important functions in the {DL}/{UL} chains. Other functions such as resource mapping/demapping, channel estimation, and scrambling are ignored. Also, the parameters such as node complexity and link weight are approximations to real-world values. With this simplification, we are able to show the essence of flexible splitting and placement of baseband functions. Still, fine-grained tuning is required if the proposed model is applied in practice.

Processing nodes are labeled based on their types, the logical cells they belong to, and the sub-chaisn they reside in. For example, the node {MIMOtx.1.2} is a {MIMO} transmitter which resides in the second {DL} processing-chain (two in total) of the first cell. Each node is assigned with a weight based on its type to reflect its computational complexity. The weight for each type is listed in Table \ref{node}. The weight value is selected based on experimental results in \cite{cloudiq}.

\begin{table}[!t]
	\renewcommand{\arraystretch}{1.3}
	\caption{Node weights with respect to node type.}
	\label{node}
	\centering
	\begin{tabular}{c||c|c|c|c|c|c}
		\hline
		\hline
		Index  &    1    &   2	   &    3	&     4	  &   5	   &   6\\
		\hline
		Type   & radioTX & radioRX &   fft  &   ifft  & MIMOtx & MIMOrx\\
		\hline
		Weight &    0    &    0    &    1   &     1   &  0.5   &  0.5   \\
		\hline
		\hline
		Index  &    7    &    8    &    9   &    10   &   11     &   12\\
		\hline
		Type   & mod     &  demod  &  code  &  decode & sourceDL & sinkUL\\
		\hline
		Weight & 0.1     & 0.1     &  0.1   &     2   & 0        &   0\\
		\hline
		\hline
	\end{tabular}
\end{table}

The weights of links are also shown in Fig. \ref{baseband1}. The magnitudes of link weights reflect the information flow between processing nodes. For example, each {MIMOrx} node gets a link from the {FFT} node with a weight of $0.45$ because we assume the overhead of cyclic prefix and control signaling is $10\%$, and the information after {CP} removal is equally divided between the two processing-chains. Also, note that the link weight is greatly increased/reduced after modulation/demodulation because we assume each $30$ bit complex baseband sample is transformed from/into a $4$ bit constellation codework ($16$-{QAM}). In case of {CoMP}, we add mutual links\footnote{The weight of {CoMP} links are equal to the links between {MIMO} and {FFT}.} between neighbouring {MIMO} modules in a cyclic fashion (e.g. {MIMOtx.n.2} - {MIMOtx.(n+1).2} and {MIMOrx.N.1} - {MIMOrx.1.1}). 

We assume computational and fronthauling cost functions to have exponential forms as shown in Table \ref{computational} and Table \ref{fronthauling}.  The computational cost at central office is zero because provisioning computational resources in central offices is less expensive. The fronthauling cost within cell sites or the data center is zero because internal information exchange does not need to use fronthaul network. The fronthauling cost between cell sites is higher than the cost between a cell site and the central office because fronthaul networks are usually optimized for centralization. Also, we assume baseband tasks in a cluster equally divides the computational resource, thus the delay of a processing functions can be represented with the product of the corresponding node weight and the total node weight in the cluster (Table \ref{delayTab}).

\begin{table}[!t]
	\renewcommand{\arraystretch}{1.3}
	\caption{Computational cost function $c_c(i,\xi)$ used in simulation.}
	\label{computational}
	\centering
	\begin{tabular}{c|c}
		\hline
		\hline
		Cell site & Central office \\
		\hline
		$2^{\sum_{\xi(v)=i}{\gamma(i)}}$    &  $0$ \\
		\hline
		\hline
	\end{tabular}
\end{table}
\begin{table}[!t]
	\renewcommand{\arraystretch}{1.3}
	\caption{Fronthauling cost function $c_f(i,j,\xi)$ used in simulation.}
	\label{fronthauling}
	\centering
	\begin{tabular}{c|c}
		\hline
		\hline
		Clusters & Cost \\
		\hline
		within cell site & 0 \\
		\hline
		within central office & 0 \\
		\hline
		between cell sites i and j & $4^{\sum_{\xi(e)=(i,j)}{\omega(e)}}$ \\
		\hline
		between cell site and central office & $2^{\sum_{\xi(e)=(i,j)}{\omega(e)}}$ \\
		\hline
		\hline
	\end{tabular}
\end{table}
\begin{table}[!t]
	\renewcommand{\arraystretch}{1.3}
	\caption{Delay function $d(p;\xi).$}
	\label{delayTab}
	\centering
	\begin{tabular}{c|c}
		\hline
		\hline
		cell site & central office \\
		\hline
		$\sum_{v \in p}{(\gamma(v)\sum_{\xi(w)=\xi(v)}\gamma(v))}$    &  $0$ \\
		\hline
		\hline
	\end{tabular}
\end{table}

\section{Graph-based Genetic Algorithm}\label{algorithm}
The clustering scheme can also be represented with a discrete valued vector $\bm{\xi} \in \mathbb{Z}^N$, where $N$ is the total number of baseband processing functions. The $k$-th entry of $\bm{\xi}$ is the cluster index for the $k$-th node. With this representation, the cost functions are parameterized by $\bm{\xi}$ and the graph clustering problem is transformed into a 2-objective combinatorial optimization problem:
\begin{equation}
\begin{array}{ll}
\min\limits_{\xi} & \sum\limits_i{c_c(i;\xi)},  \text{ } \sum\limits_{i}\sum\limits_{j}{c_f(i,j;\xi))}\\
s.t. & d(p;\xi)=\sum_{v\in p}{d(v,p;\xi)}<D(p).
\end{array}
\end{equation}
It is difficult to give a general analytical solution to such a problem. So we turn to genetic algorithm (GA) to find sub-optimal solutions. The basic building blocks of GA are selection, crossover, and mutation. A typical GA session is initialized with a population carrying a collection of genes. The algorithm then iteratively loop through the three basic building blocks until solution converges or some termination conditions are met. From the perspective of computation, GA can also be seen as embedded parallel algorithms which search for the ``good'' solution by simultaneously experimenting with multiple solutions. Although GA can be applied to many types of problems, its performance will become satisfactory only after some  customization. For this reason, we designed a customized GA to solve the graph-clustering problems described above.

\subsection{Natural encoding and cluster seeding}
A key problem with GA is how to represent solutions as combinations of genes (chromosome). This process is also called encoding. Good encoding should make it easy to produce legitimate offspring individuals through crossover and mutation. Here we directly use the clustering vector $\bm{\xi}$ for encoding. The advantage of this encoding scheme should be obvious when we present our crossover and mutation functions. Notice that we keep some nodes in a fixed cluster to reflect the fact that some functions can only be place at specific locations\footnote{Radio transmitter or receiver have to be placed at distributed cell sites. Assigning them to other physical locations does not make sense.}. Hereafter we refer to these nodes as ``seed nodes''. We name them in this way because the whole clustering scheme is generated based on the initial cluster assignments of seed nodes.

\subsection{Linearly combined fitness function}
Another important aspect is how to evaluate solutions with a fitness function. This problem is complicated because we have two (possibly contradicting) optimization objectives. To achieve different tradeoffs between these costs, we linearly combine computational and fronthauling cost to form a single cost function. Also, we have to incorporate the path delay constraints. Yet explicitly examining whether a solution violates these constrains makes crossover and mutation difficult. Hence we implicitly incorporate the path delay constraints as a ``penalty function'', which will significantly degrade the fitness of a solution if the path delay constraint is violated. Summing up, the overall fitness function is as follows:
\begin{equation}\label{costFun}
\begin{aligned}
F(\bm{\xi};\alpha,\beta) =&{ } \alpha \sum_i{c_c(i;\xi)}
						  + (1-\alpha) \sum_{i}\sum_{j}{c_f(i,j;\xi))}\\
						  &+ \beta \sum_{p}(d(p;\xi)-D(p))^+,
\end{aligned}
\end{equation}
where $0 \le \alpha \le 1$ is the tradeoff coefficient, $\beta > 1$ is the penalty coefficient, and $(\cdot)^+$ is the non-negative clipping function.
\subsection{Dispersive crossover}
The crossover function we choose is dispersive crossover. This crossover function selectes the genes of an offspring from its parents with equal probability. With natural encoding, we are guaranteed that the offspring of legitimate parents is naturally legitimate.

\subsection{Graph-based mutation}
Mutation function helps the population's chromosomes escape from local minima. Based on the structure of our problem, we tailored a customized mutation function called \emph{graph-based mutation}. We first define the connection matrix $\bm{C}$, the entries of which take on values of either $1$ or $0$. $\bm{C}(i,j) = 1$ if and only if node $i$ and node $j$ are connected. Using $\bm{C}$, we can define the \emph{Allowed Mutation Set} as $\bm{A}(i) = \{\xi(j) \mid \bm{C}(i,j) = 1, j \text{ is seed}\},$ which gives all the clusters that a node $i$ is currently connected with. In summary, graph-based mutation can be described as follows: we randomly change the value of an individual's chromosome at position $i$ to a value selected from $\bm{A}(i)$. In this way, we can avoid ``bad'' mutations because placing a node to a unconnected cluster will only increase the total cost.

\section{Simulation Results}\label{simulation}
In this section, we apply the customized genetic algorithm and discuss simulation results. We use the parameters of the example in \ref{Example} in our simulation. To avoid small dynamic range of $\alpha$, we rescale the computational and fronthauling cost with respect to their maximum value. Other important parameters of {GA} are shown in Table \ref{algo}. Note that the initialization function is also graph-based, i.e. we intialize nodes to the clusters that have connected seeds.

\begin{table}[!t]
	\renewcommand{\arraystretch}{1.3}
	\caption{Algorithm parameters.}
	\label{algo}
	\centering
	\begin{tabular}{c|c}
		\hline
		\hline
		Parameter                   &     Value / Type \\
		\hline		
		Population size             &     $20$ \\
		\hline
		Initialization              &     Graph-based random initialization  \\
		\hline
		Seeds              			&     RadioTx, RadioRx, SourceDL, SinkUL   \\		
		\hline
		Selection             		&     Rolling-wheel selection \\
		\hline
		Crossover                   &     Dispersive crossover  \\		
		\hline
		Mutation                    &     Graph-based mutation (Prob. $= 0.4$)   \\
		\hline
		Delay penalty factor		&     $10$\\
		\hline
		\hline
		\end{tabular}
		\end{table}

\subsection{Tradeoff between computation and fronthauling costs}
\begin{figure}[!t]
\centering
\includegraphics[width=3.2in]{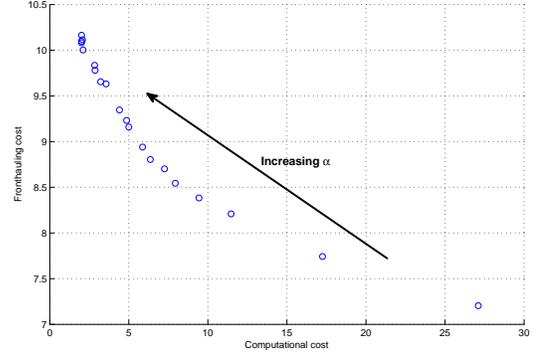}
\caption{Tradeoff between two costs by varying $\alpha$, $D(p)=30$.}
\label{tradeoff}
\end{figure}

\begin{figure}[!t]
	\centering
	\includegraphics[width=3.2in]{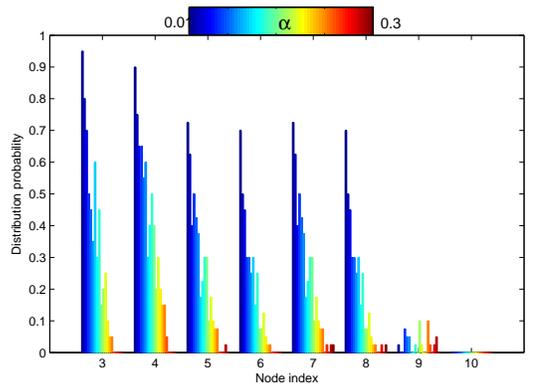}
	\caption{Clustering schemes under different $\alpha$, $D(p)=30$.}
	\label{tradeoffProfile}
\end{figure}

Next we show how the proposed algorithm can achieve different tradeoffs between computational and fronthauling costs by varying parameter $\alpha$. Fig. \ref{tradeoff} shows the average (over 10 simulation runs) computational and fronthauling costs using $\alpha \in [0.01, 0.3]$. The tradeoff between these two costs can be clearly observed: when $\alpha$ increases, computational cost is reduced while fronthauling cost is increased. To understand how this tradeoff is possible, we show the corresponding clustering schemes in Fig. \ref{tradeoffProfile}. The x-axis indicates the node indexes, while the y-axis indicates the probability that a certain type of node is distributed at cell sites\footnote{We do not show distributions for radioTX, radioRX, sourceDL, and sinkUL because they are seed nodes and have deterministic cluster index.}. As $\alpha$ increases (color becomes warmer), the computational cost of placing processing functions at remote sites also increases. As a result, more nodes are centralized to the central office to save computational resources at the expense of increased fronthauling cost. Also notice that, no matter what value $\alpha$ takes on, the decode nodes are always centralized. This is because these nodes has very high computational complexity, the schemes which place them at remote sites has large delay penalty and is prohibited. This phenomenon provides an intuitive guideline to centralize computation-intensive functions.

\subsection{Influence of cooperative structures}
\begin{figure}[!t]
\centering
\includegraphics[width=3in]{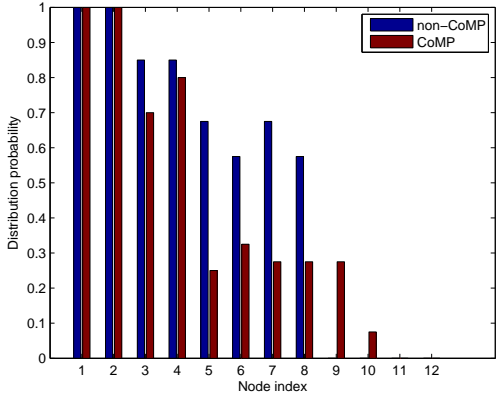}
\caption{Clustering statistics with and without {CoMP}, $\alpha=0.05$, $D(p)=30.$}
\label{comp}
\end{figure}

The presence of cooperative processing structures has great influence on the outcomes of our algorithm. Here we compare the simulation results of baseband structures with and without {CoMP}. The clustering statistics (averaged over $10$ simulation runs) are shown in Fig. \ref{comp}. As can be seen, more baseband function are centralized under the presence of {CoMP} compared with non-{CoMP} scenario. The reason is that, cooperating {MIMO} functions have large interconnection bandwidth requirements, but the fronthauling cost between distributed cell sites is high. Distributed placement of {CoMP} modules will incur high fronthauling cost and should thus be avoided. In real networks, not all resource blocks are scheduled for {CoMP} operation. In that case, we can combine the results of {CoMP} and {non-CoMP} cases to save fronthaul bandwidth. Specifically, we can centralize only cooperating {MIMO} functions, and leave other {MIMO} funcitons at cell sites. In this way, the links between cell sites and central office will have less total bandwidth.

\subsection{Influence of delay constraints}
\begin{figure}[!t]
\centering
\includegraphics[width=3in]{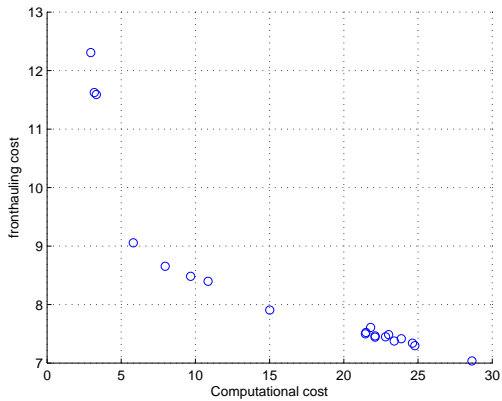}
\caption{Clustering statistics with different delay constraint, $\alpha=0.01,$ w/o {CoMP}.}
\label{delay}
\end{figure}

We also investigate the influence of delay constraint. In Fig. \ref{delay}, we show the average (over 10 simulation runs) computational and fronthauling cost under delay thresholds ranging from $1$ to $20$. As can be observed in this figure, different delay threshold will result in different tradeoff. Smaller threshold values make distributed placement prone to higher delay penalty, thus the resulting clustering scheme favors centralization and have higher fronthauling cost. In contrast, we can place more functions at remote sites when the delay bounds get looser.

\section{Conclusion}\label{conclusion}
In this paper, we present a graph-based framework for analyzing baseband function splitting and placement problems in {C-RAN}. We re-express baseband processing structures with a graph model, and transform splitting and placement strategies into graph-clustering schemes. To solve the desired tradeoffs between computational and fronthauling costs, we present a genetic algorithm with customized fitness function and mutation module. Simulation results show that we can achieve arbitrary cost tradeoffs by varying algorithm parameter. The investigation on {CoMP} and delay constraint also give important implications for function splitting in realistic systems. As a future work, we plan to apply the proposed framework to other baseband structures and investigate the cost tradeoff characteristics of these structures. Also, we plan to further tune the graph and algorithm parameters according to realistic baseband processing structures so that the results can be more practical. 

\section*{Acknowledgment}
This  work  is  sponsored  in  part  by  the  {National Basic  Research  Program  of China  (973 Program: 2012CB316001}), {the National Science Foundation of China (NSFC) under grant No. 61201191 and No. 61401250}, {the Creative Research Groups of NSFC under grant No. 61321061}, and {Intel Collaborative Research Institute for Mobile Networking and Computing}.


\bibliographystyle{IEEEtran}
\bibliography{IEEEabrv,myref}

\end{document}